\documentclass[aps,prl,groupedaddress,twocolumn,10pt]{revtex4-2}

\usepackage{amsmath}
\usepackage{amsfonts}
\usepackage{amssymb}
\usepackage{times}
\usepackage{mathtools}
\usepackage{circuitikz}
\tikzset{every picture/.style={line width=0.2mm}}
\ctikzset{bipoles/thickness=1,voltage=straight}

\newcommand{\kb}{k_{\mathrm{B}}} \newcommand{\dr}{\partial} \newcommand{\ee}{\mathrm{e}} \newcommand{\dd}{\mathrm{d}}     \newcommand{\llangle}{\langle\!\langle} \newcommand{\rrangle}{\rangle\!\rangle}
\DeclareMathOperator{\sgn}{\mathrm{sgn}}

\newenvironment{eqal}{\begin{equation}\begin{aligned}}{\end{aligned}\end{equation}\ignorespacesafterend}

\newcommand{\eq}{\textrm{Eq.\,}} \newcommand{\fig}{\textrm{Fig.\,}}

\begin{document}

\title{Environmental Effects and Brillouin's Paradox in Nonlinear Electrical Circuits}

\author{Lucas Désoppi}
\email[]{lucas.desoppi@usherbrooke.ca}
\affiliation{Département de physique, Institut quantique, Université de Sherbrooke, Sherbrooke, Québec, Canada J1K 2R1}

\author{Bertrand Reulet}
\affiliation{Département de physique, Institut quantique, Université de Sherbrooke, Sherbrooke, Québec, Canada J1K 2R1}


\date{\today}

\begin{abstract}
We present a stochastic approach to calculate the full statistics of classical voltage fluctuations across an arbitrary, nonlinear, dissipative device embedded in a circuit in the presence of a DC bias. We show how the feedback resulting from the circuit, made of an ohmic resistor and a capacitor, affects the statistics of voltage fluctuations, and in particular resolves Brillouin's paradox to satisfy thermodynamics: feedback cancels rectification. We apply our very general results to the case of a tunnel junction and a diode.
\end{abstract}


\maketitle

What happens when two electronic components are connected in series in a circuit ? This naive question refers to the basis of electronic circuit design: each component has its own, intrinsic $I_{1,2}(V)$ characteristic, and the voltage drops $V_{1,2}$ across the components 1,2 adjust themselves so that: i) the DC current in the circuit $I$ is the same in the components, $I=I_1(V_1)=I_2(V_2)$, and ii) the total voltage is $V=V_1+V_2$. This is however known to be incorrect in mesoscopic circuits at very low temperature, a phenomenon called dynamical Coulomb blockade \cite{Fulton87,Delsing89,Korotkov90,Averin91}. This breakdown of the most basic electronic formula has been recently shown to be much more general: it occurs for any circuit with a device whose noise depends on voltage, like an avalanche diode in series with a resistor, at room temperature \cite{Karl21}.  The physical origin of this lies in the feedback of current noise: when a component generates noise whose amplitude depends on voltage, the presence of a resistor (or more generally any complex impedance) in series with itself leads to a self-modulation of noise, which results in the modification of the I-V characteristic. This can be strong enough to induce negative differential resistance in a Zener diode, making it possible to build an amplifier with just a Zener diode and a resistor, something  impossible in "usual" electronics\cite{Dumont26}. 

The effect of noise feedback on the DC characteristic of a tunnel junction has been thoroughly calculated in a quantum framework both analytically and numerically at zero temperature \cite{Devoret90, Ingold1992, Zaikin05, Hakonen06, Zaikin12, Souquet13, Grabert17}. The tunnel junction has a linear I-V characteristic and exhibits shot noise. In \cite{Karl21} a more general classical treatment is applied to calculate the average current and noise, supposed Gaussian. The approach allows to calculate the effect of a small resistor in series with any component and proceeds recursively. It however suffers from not considering the circuit dynamics, which results in problems with causality and does not consider the effect of finite bandwidth. In the present article, we include the dynamics of the circuit and treat on the same footing all cumulants of voltage/current fluctuations.

Our approach, based on the theory of diffusive stochastic processes, allows for a calculation of the stationary probability density of voltage fluctuations. We deduce the first three cumulants for any noisy, nonlinear component and discuss their link with thermodynamics. In particular, we show how feedback effects resolve the Brillouin paradox, i.e. that a nonlinear component might rectify its own noise to generate a DC voltage, in contradiction with the first principle of thermodynamics \cite{Brillouin50}. After extending our approach to the case of jump processes, we apply our results to the case of the tunnel junction and the diode. In both cases we obtain a correction to DC transport related to the presence of the Coulomb gap, usually obtained by quantum mechanical theories.

\begin{figure}
    \centering
    \begin{circuitikz}[scale=1.0]
    \draw (0,0) -- (0,0.8);
    \draw (-0.7,0.8) -- (0.7,0.8) to[C,l=$ C $] (0.7,2.5) -- (-0.7,2.5) to[generic, v<=$ U_t $] (-0.7,0.8);
    \node at (-0.7,1.65) {$I^{\textsc{d}}_t$};
    \draw (0,2.5) -- (0,3.3) -- (2, 3.3);
    \draw (0,0) -- (2,0) to[vsource, v_= $ V $] (2,1.65);
    \draw (2,1.65) to[R, l_=$R$] (2,3.3);
\end{circuitikz}
    \caption{Schematics of the circuit.}
    \label{fig:Circuit}
\end{figure}
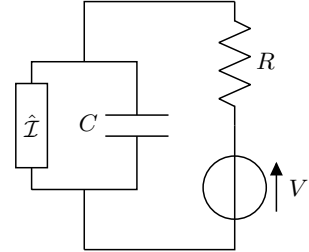

\paragraph*{Electrical circuit model.} In this letter, we consider the simplest circuit that embeds a nonlinear, noisy device, with feedback provided by the resistor $R$ at temperature $ T_R $ and dynamics due to the capacitor $C$, driven out-of-equilibrium by a voltage source $V$, see \fig \ref{fig:Circuit}. This circuit obeys:
\begin{eqal}
    C \dot{U}_t & = (V-U_t)/R - \sqrt{2\mathcal{D}_{\textsc{r}}} \, \dot{W}^{\textsc{r}}_t - I_{t}^{\textsc{d}},
    \label{EOM}
\end{eqal}
$ W^{\textsc{r}}_t $ is a standard Wiener process that models the Johnson noise of the feedback resistor, of noise spectral density $2\mathcal{D}_{\textsc{r}}=2\kb T_{\textsc{r}}/R$ and  $ I_{t}^{\textsc{d}} $ represents the instantaneous current through the noisy nonlinear device and $U_t$ the voltage across it.
We do not work directly with the stochastic equation of motion \eqref{EOM}, but write the corresponding master equation for the probability density function $ P_t(u) $ of measuring $ U_t = u $ at time $t$, which takes the form:
\begin{eqal}
    \dot{P}_t(u) = \dfrac{\dr }{\dr u} \bigg[ \dfrac{u-V}{R C} P_t(u) + \dfrac{\kb T_{\textsc{r}}}{R C^2} \dfrac{\dr P_t}{\dr u}(u) \bigg]  + \hat{\mathcal{L}} P_t(u),
    \label{tMeq}
\end{eqal}
where $ \hat{\mathcal{L}} $ is a linear operator acting on $ P_t(u) $, whose expression is equivalent to $  I_{t}^{\textsc{d}}   $ (as seen in the following examples).

\paragraph*{Diffusive noise models.} We need to specify what the operator $ \hat{\mathcal{L}} $ (or equivalently $ I_{t}^{\textsc{d}} $) is. Diffusive noise models constitute the privileged class of models considered in this work, for which one can write:
\begin{eqal}
     I_{t}^{\textsc{d}} = \mathcal{I}(U_t) + \sqrt{2 \mathcal{D}(U_t)} \bullet \dot{W}_t,
\end{eqal}
where $W_t$ is a standard Wiener process, $ \mathcal{I}(u) $ is the deterministic I-V characteristic, $ \mathcal{D}(u) $ is the voltage-dependent current noise spectral density (in $ \mathrm{A}^2 \cdot \mathrm{Hz}^{-1} $), supposed frequency independent. The bullet $ \bullet $ means that the stochastic differential equation \eqref{EOM} is to be interpreted according to the Hänggi-Klimontovich prescription (see discussion below), for which the operator $ \hat{\mathcal{L}} $ is given by:
\begin{eqal}
     \hat{\mathcal{L}} P(u) =   \dfrac{\dr }{\dr u} \bigg( \dfrac{\mathcal{I}(u)}{C} P(u) + \dfrac{\mathcal{D}(u)}{C^2} \dfrac{\dr P}{\dr u}(u)  \bigg).
     \label{DiffOp}
\end{eqal}
In the stationary state, the probability density $ P_{\mathrm{st}}(u) $ obeys:
\begin{eqal}
    \dfrac{\dd }{\dd u} \bigg[ \dfrac{u-V}{R C} P_{\mathrm{st}}(u) + \dfrac{\kb T_R}{R C^2} \dfrac{\dd P_{\mathrm{st}}}{\dd u}(u) \bigg]  + \hat{\mathcal{L}} P_{\mathrm{st}}(u) = 0.
    \label{StMEq}
\end{eqal}
Taking into account the boundary conditions $ P_{\mathrm{st}}(u\to \pm \infty) = 0 $, we obtain
\begin{eqal}
    P_{\mathrm{st}}(u) & = \ee^{- \mathcal{A}(u)}/\mathcal{Z}, \\
     \mathcal{A}(u) & = C \int_{u_0}^u  \dd u' \, \dfrac{\mathcal{I}(u') + (u'-V)/R}{\mathcal{D}(u') + \mathcal{D}_{\textsc{r}}},
     \label{StDistrib}
\end{eqal}
where $\mathcal{Z}$ is a normalization constant and $u_0$ an arbitrarily chosen voltage. It is useful to define the fluctuation variable $\tilde{U}$ so that $U = u_0 + \tilde{U}$, and to choose the most probable voltage for $u_0$, satisfying $ \mathcal{A}'(u_0) = 0 $ and $ \mathcal{A}''(u_0) > 0 $. The former condition explicitly gives:
\begin{eqal}
    R \mathcal{I}(u_0) +u_0 = V ,
    \label{IdealIV}
\end{eqal}
thus $u_0$ is also the voltage across the nonlinear device in the absence of noise. This results from the choice of the Hänggi-Klimontovich prescription \cite{Desoppi25,Bonnin25}.

Writing $ u = u_0 + \tilde{u} $, the expression of the 'action' $\mathcal{A}(u) $ near $ u_0 $ takes the form
\begin{eqal}
    \mathcal{A}(u) = C \int_0^{\tilde{u}} \dd \tilde{u}' \dfrac{\tilde{u}'/R + \mathcal{I}(u_0 +\tilde{u}') - \mathcal{I}(u_0)   }{ \mathcal{D}(u_0 +\tilde{u}') + \mathcal{D}_{\textsc{r}}},
\end{eqal}
The numerator is the sum of the fluctuation $\tilde{u}'/R$ and the feedback term $ \mathcal{I}(u_0+\tilde{u}') - \mathcal{I}(u_0)  $. The denominator is the total noise $\mathcal{D}_\mathrm{tot}(u)=\mathcal{D}(u)+\mathcal{D}_{\textsc{r}}$. If the device is an ohmic resistor and $T_{\textsc{r}}=T$, we find $\mathcal{A}(u)=Cu^2/(2\kb T)$ as expected for RC circuit at temperature $T$. The same result is recovered for an arbitrary device shunted by a small resistor $R\to 0$, or for an arbitrary $R$ with a device that obeys $\mathcal{D}(u)=\kb T\mathcal{I}(u)/u$ \cite{Desoppi25}.

\paragraph*{Perturbative method.} From \eq \eqref{StDistrib}, we could compute numerically the moments of the voltage $ \langle U^n \rangle $, however, in this work, we focus on analytical computations using perturbation theory. The action can be decomposed as a sum of a quadratic term and a perturbation:
\begin{eqal}
    \mathcal{A}(u_0+\tilde{u}) & = \mathcal{A}''(u_0) \dfrac{\tilde{u}^2}{2} + \delta\mathcal{A}(\tilde{u}),  \\
    \mathcal{A}''(u_0)&  =  \dfrac{C \big(1+R \mathcal{I}'(u_0)\big)}{R\mathcal{D}_\mathrm{tot}(u_0)}, \\
    \delta\mathcal{A}(\tilde{u}) & = \mathcal{A}^{(3)}(u_0)\dfrac{ \tilde{u}^3}{3!} + ...
\end{eqal}
$\mathcal{A}''(u_0)^{-1}$ is the variance of voltage fluctuations of a current noise source of spectral density $S_\mathrm{tot}=2\mathcal{D}_\mathrm{tot}(u_0)$ into the parallel combination $R_\parallel=R(1+RG)^{-1}$ of the resistor $R$ and the device of differential conductance $G=\mathcal{I}'(u_0)$, integrated over the total noise bandwidth $B=(2R_\parallel C)^{-1}$: $\mathcal{A}''(u_0)^{-1}=R_\parallel^2S_\mathrm{tot}B$. 

A truncation at order $3$ in $\tilde{u}$ will be sufficient for what follows. Indeed, the feedback effects manifest themselves because  $\mathcal{A}^{(3)}(u_0)\neq0 $ in general. We find:
\begin{eqal}
    \mathcal{A}^{(3)}(u_0) 
    & = \dfrac{C\mathcal{I}''(u_0)}{\mathcal{D}_\mathrm{tot}(u_0)} - 2  \dfrac{ \mathcal{D}'(u_0)\mathcal{A}''(u_0)}{\mathcal{D}_\mathrm{tot}(u_0)}.
\end{eqal}
The deviation from a Gaussian distribution stems from the nonlinearity of the I-V characteristic, i.e. $\mathcal{I}''\neq0$, or the voltage dependence of the noise, $\mathcal{D}'\neq0$. Noting the Gaussian average $
    \langle f(\tilde{U})\rangle_{\mathrm{G}} = \sqrt{\mathcal{A}''(u_0) / 2\pi} \int_{\mathbb{R}} \dd \tilde{u} \, f(\tilde{u}) \, \ee^{ -\mathcal{A}''(u_0) \tilde{u}^2/2}$, 
with which $    \langle f(U) \rangle =  \big\langle f(u_0+\tilde{U}) \ee^{ -\delta\mathcal{A}(\tilde{U})}   \big\rangle_{\mathrm{G}}  \big/ \big\langle \ee^{ -\delta\mathcal{A}(\tilde{U}) }  \big\rangle_{\mathrm{G}} $, a Taylor expansion of the exponential functions containing the perturbation $ \delta\mathcal{A}(\tilde{U}) $ is performed, up to the desired order. This expansion is valid as long as $ |\delta\mathcal{A}(\tilde{u})| \ll \mathcal{A}''(u_0) \tilde{u}^2/2 $ for $|\tilde{u}|$ not exceeding a few standard deviations (typically $ |\tilde{u}| \lesssim 3 /\sqrt{\mathcal{A}''(u_0)}  $).

The first order in the perturbation $ \delta\mathcal{A}(\tilde{U}) $, gives (with $n$ an integer number):
\begin{equation}
    \langle \tilde{U}^n \rangle  \approx  \langle \tilde{U}^n \rangle_{\mathrm{G}} + \langle \tilde{U}^n \rangle_{\mathrm{G}} \langle \delta\mathcal{A}(\tilde{U}) \rangle_{\mathrm{G}} - \langle \tilde{U}^n \delta\mathcal{A}(\tilde{U}) \rangle_{\mathrm{G}} .
\end{equation}
Taking into account solely the cubic term in the perturbation, and making use of the following standard properties associated to the Gaussian integral:
\begin{eqal}
    \langle \tilde{U}^{2 n+1} \rangle_{\mathrm{G}} = 0, \quad \langle \tilde{U}^{2 n} \rangle_{\mathrm{G}} = \dfrac{(2 n)!}{n!\big[2\mathcal{A}''(u_0) \big]^n},
\end{eqal}
one can calculate all the cumulants of $\tilde{U}$. 

The correction to the DC voltage reads:
\begin{equation}
     \langle \tilde{U} \rangle \approx 2 R_\parallel^2 \mathcal{D}'(u_0) B
     - R_\parallel^3 \mathcal{D}_\mathrm{tot}(u_0)\mathcal{I}''(u_0) B.
     \label{eq:U_average}
\end{equation}
The first term corresponds to the usual dynamical Coulomb blockade, that can be formulated as the integral over frequency of $\frac12\chi(\omega)\mathrm{Re}Z(\omega)$ with $\chi$ the noise susceptibility \cite{Gabelli08}. Here $\chi=2\mathcal D'$, $Z=R_\parallel$ and the integration over frequency gives the bandwidth $B$. The usual calculation of dynamical Coulomb blockade considers $Z$ to be the environmental impedance only, not including the device. In contrast, we find the parallel combination of $R$ and the device. This comes from the usual hypothesis that the sample, a tunnel junction, is much more resistive than the environment, a hypothesis we do not need here. The second term in $\langle \tilde{U}\rangle$ corresponds to the rectification of the total voltage noise generated by the device and the feedback resistor. It is interesting to consider the equilibrium $V=u_0=0$, $T_{\textsc{r}}=T$. Finding $\langle \tilde{U}\rangle=0$ imposes:
\begin{equation}
    2\mathcal{D}'(0)=\kb T\mathcal{I}''(0)
    \label{eq:FDT}
\end{equation}
which is a particular form of the fluctuation dissipation theorem: in the same way as the equilibrium noise is proportional to the conductivity, the equilibrium noise susceptibility is proportional to the nonlinear conductance \cite{Efremov68}. Thus at equilibrium the phenomenon of feedback exactly compensates the rectification of the noise: the resolution of Brillouin's paradox \cite{Brillouin50,vanKampen60,Gunn68,Buttiker12} comes from the feedback. For $T\neq T_{\textsc{r}}$ and using \eq \eqref{eq:FDT} we find:
\begin{eqal}
    \langle \tilde{U} \rangle = \kb (T - T_{\textsc{r}}) R_{\parallel}^3 R^{-1} B \mathcal{I}''(0), 
\end{eqal}
and then the current is given by $ \langle I\rangle = - \langle \tilde{U}\rangle/R $. This expression is qualitatively in agreement with the results from \cite{Gunn68, Gunn69}, even though the factor  $ R^3_\parallel/R^2 $ is different, thereby motivating a  quantitative experimental study. Let us note that our Eq.(\ref{eq:U_average}) allows us to calculate the combined effects of a temperature difference and a voltage bias.

The variance of voltage fluctuations reads:
\begin{equation}
 \llangle U^2 \rrangle \approx 2 R_\parallel^2 \mathcal{D}_\mathrm{tot}(u_0) B  . 
 \label{eq:U2}
\end{equation}
There is no feedback correction on this term to first order in $\delta\mathcal{A}$. However, the noise for a given average voltage $\langle U\rangle$ will be the one obtained in the absence of feedback for a voltage $u_0$. At equilibrium, we find $\llangle U^2 \rrangle = \kb T/C$ as expected from thermodynamical considerations.

Using $ \llangle I^2 \rrangle = \llangle U^2 \rrangle/R^2 $, we find, in the approximation of a constant bandwidth $B$ and a small resistance $R$ whose noise is neglected:
\begin{eqal}
    \dfrac{\dr \langle I \rangle}{\dr R} & \approx - \langle I \rangle \dfrac{\dr \langle I \rangle}{\dr V} - \dfrac{\dr \llangle I^2 \rrangle}{\dr V}, \\
    \dfrac{\dr \llangle I^2 \rrangle}{\dr R} & \approx - 2 \llangle I^2 \rrangle \dfrac{\dr \langle I \rangle}{\dr V} - \langle I \rangle \dfrac{\dr \llangle I^2 \rrangle}{\dr V}. 
\end{eqal}
These expressions are in close correspondence to those derived in \cite{Karl21}. They differ only by a factor $ 1/2 $ in front of $ \dr \llangle I^2 \rrangle/\dr V $, because the Stratonovich convention was implicitly chosen.

The skewness of $U$ reads:
\begin{eqal}
    \llangle U^3 \rrangle & \approx
     8 R_\parallel^4 \mathcal{D}_\mathrm{tot}(u_0) \mathcal{D}'(u_0) B^2 \\
    & \qquad - 4 R_\parallel^5\mathcal{D}_\mathrm{tot}^2(u_0)\mathcal{I}''(u_0)B^2.
    \label{eq:U3}
\end{eqal}
The first term corresponds to the usual environmental correction established for a linear conductor \cite{Beenakker03,Reulet03,Kindermann04}, where the spectral density of the third cumulant of voltage fluctuations is found to be $3S_{VV} \dd S_{VV}/\dd V$ with $S_{VV}=2\mathcal{D}_\mathrm{tot} R_\parallel^2$ the voltage noise spectral density. The second term arises from the nonlinearity of the device: a Gaussian current noise with spectral density $S$ applied to a nonlinear device obeying $U=R_\parallel I +\alpha I^2$ leads to voltage fluctuations with a skewness of spectral density $3\alpha R_\parallel^2S^2$ with $\alpha=-R_\parallel^3 \mathcal{I}''(u_0)/2$. The expression of the skewness can be put in a compact form:
\begin{equation}
    \llangle U^3\rrangle=2 \langle\tilde U\rangle\llangle U^2\rrangle.
\end{equation}
This relation highlights the profound link between the environmental corrections to the skewness and to the DC voltage, usually ascribed to dynamical Coulomb blockade. At equilibrium, the environmental contribution to the skewness vanishes. However, a finite skewness is predicted at zero voltage bias if $T\neq T_{\textsc{r}}$.

\paragraph*{Continuous approximation for jump models.} Jump noise models constitute another important class of models, for which, as mentioned in the introduction, exploiting \eq \eqref{StMEq} is generally difficult.  For concreteness, we will concentrate on random jumps of one electron at a time, in which case two characteristic voltages appear, given by $ \kb T/e $ and $  e/C $, where $ e > 0 $ is the elementary electric charge. When the continuous limit (that is, the Fokker-Planck approximation) of these processes is accurate, as described in \cite{Desoppi25}, one can use the results of the previous sections. This limit is obtained by mapping the transition rates on the former diffusive model, according to the expressions
\begin{eqal}
    2 \mathcal{D}(u) & = e \mathcal{I}(u) \coth(e \, u/2\kb T), \\
    &  \approx 2 \kb T \,  \mathcal{I}(u)/u \hspace*{4mm} \text{(high $T$),} \\
    & \approx  e | \mathcal{I}(u)|  \hspace*{13mm} \text{(low $T$).} 
    \label{eq:shot_noise}
\end{eqal}
In the high temperature limit $ | e \, u | \ll \kb T $, we recover the generalized Johnson-Nyquist relation \cite{Desoppi25}, while the low temperature regime corresponds to pure shot noise \cite{Blanter00,Gillespie00}. The continuous limit is expected to be accurate when $ |\langle U \rangle| \gtrsim e/C $, that is, when the jump size is sufficiently small, then the "voltage ladder" given by integer multiples of $ e/C $ appears to be continuous. One can qualitatively understand this with the Central Limit Theorem: Since the jumps are approximately independent and identically distributed, a Gaussian approximation should be accurate when a sufficiently large number of jumps add up.

\begin{figure}
    \centering
    \includegraphics[width=\linewidth]{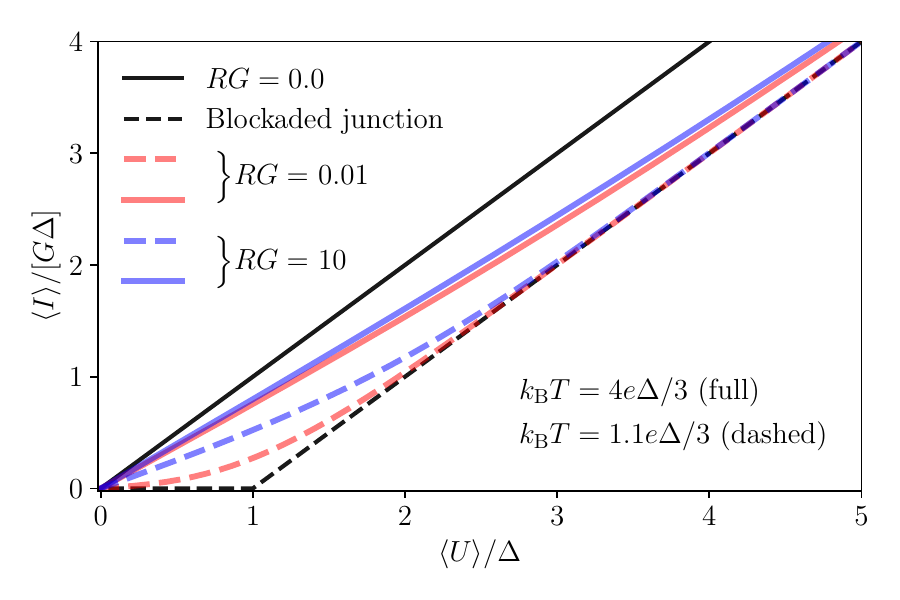}
    \caption{Rescaled I-V characteristics of a tunnel junction. The full black line corresponds to the ideal I-V characteristic $\langle I\rangle=G\langle U\rangle$, the dashed black line to the fully blockaded junction and the colored lines correspond different values of $RG$. }
    \label{fig:I-V_TJ}
\end{figure}

\paragraph*{Tunnel junction.}
A tunnel junction is a device with linear I-V characteristic, $ \mathcal{I}(u) = G u $, and with noise given by \eq\eqref{eq:shot_noise}. In this case, \eq \eqref{IdealIV} gives $ u_0 = V/(1+RG) $.  Because $ \mathcal{I}''(u) = 0 $, the rectification-induced term in $ \langle \tilde{U} \rangle $ vanishes. In the limit of a large voltage bias $V$ compared to $ \kb T/e $, one gets:
\begin{eqal}
    \langle \tilde{U} \rangle & =  R_{\parallel} G \Delta\sgn(V),
\end{eqal}
which in turn gives:
\begin{eqal}
    \langle I \rangle = G \Big(  \langle U \rangle - \Delta\sgn\big( \langle U \rangle \big)  \Big).
\end{eqal}
with $\Delta=e/(2C)$ the usual Coulomb gap. This result is identical to that of the quantum calculation \cite{Ingold1992}. In particular, the result is independent of the value of $R$ as long as $R\neq0$. The validity criterion of the perturbative expansion can be rewritten in the form $   |\langle 2\tilde{U}\rangle| \leqslant \sqrt{\llangle U^2 \rrangle}$, which is obeyed in the results presented above.
A problem however occurs at small environmental resistor $R$: one naively expects the feedback corrections to vanish as $R\to 0$, whereas our calculation shows the contrary, see dashed lines in \fig \ref{fig:I-V_TJ}. This comes from the bandwidth diverging in this limit, while the current noise of the device has unlimited bandwidth. It results in the variance of voltage fluctuations  $\llangle U^2\rrangle$ remaining finite as $R\to0$, while it is strictly zero if $R=0$ (perfect voltage bias). In a more realistic circuit the bandwidth will be limited by parasitic components such as the geometrical inductance of $R$ or of the wires between the components, so $B$ will remain finite and the feedback correction will tend to zero as $R\to0$. In the quantum realm \cite{Devoret90, Flensberg92,Safi04,Souquet13}, the bandwidth is also limited by the voltage across the device, $B\simeq e\langle U\rangle/h$. This restricts our calculation to large enough voltage, $|\langle U\rangle| >g_\parallel\Delta$ with $g_\parallel=R_Q/R_\parallel$ and $R_Q=h/e^2$ the quantum of resistance.

We show in \fig \ref{fig:I-V_TJ} the rescaled I-V characteristics of a tunnel junction at moderate temperature for various values of the product $RG$. There is clearly a well defined limit for $R \to\infty$, since the amount of fluctuations seen by the device depends on $R_\parallel$, not $R$ alone. As soon as $T>\Delta$, the result is remarkably almost independent on $RG$, see full lines in \fig \ref{fig:I-V_TJ}.

\begin{figure}
    \centering
    \includegraphics[width=1.0\linewidth]{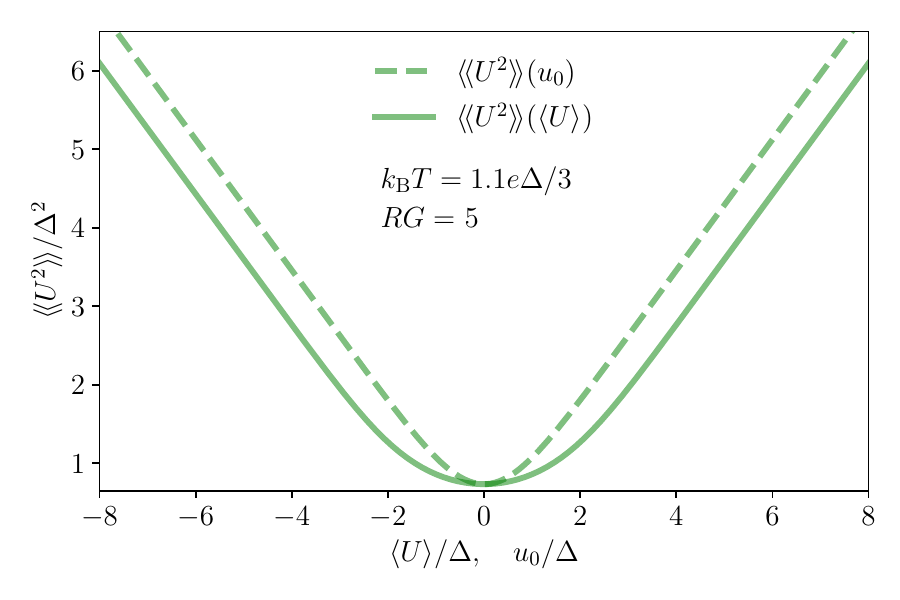}
    \caption{Rescaled variance of voltage fluctuations in a tunnel junction, $ \llangle U^2\rrangle/\Delta^2 $ as a function of $u_0/\Delta$ (dashed line) and $\langle U\rangle /\Delta$ (solid line) for $RG=5$ and $\kb T=1.1e\Delta/3$. The value at equilibrium (zero bias voltage) corresponds to $\llangle U^2\rrangle=\kb T/C$ for both curves.} 
    \label{fig:U2_TJ}
\end{figure}

The voltage noise of the junction in the presence of the external resistor $R$ is given by Eqs. (\ref{eq:U2}) and (\ref{eq:shot_noise}) taken at the voltage $u_0=\langle U\rangle-\langle \tilde U\rangle$. The result is shown in \fig \ref{fig:U2_TJ}, where the rescaled voltage noise variance is plotted as a function of $u_0$ (dashed line), which corresponds to the absence of feedback effects, and as a function of the real average voltage $\langle U\rangle$ (solid line).

\begin{figure}
    \centering
    \includegraphics[width=1.0\linewidth]{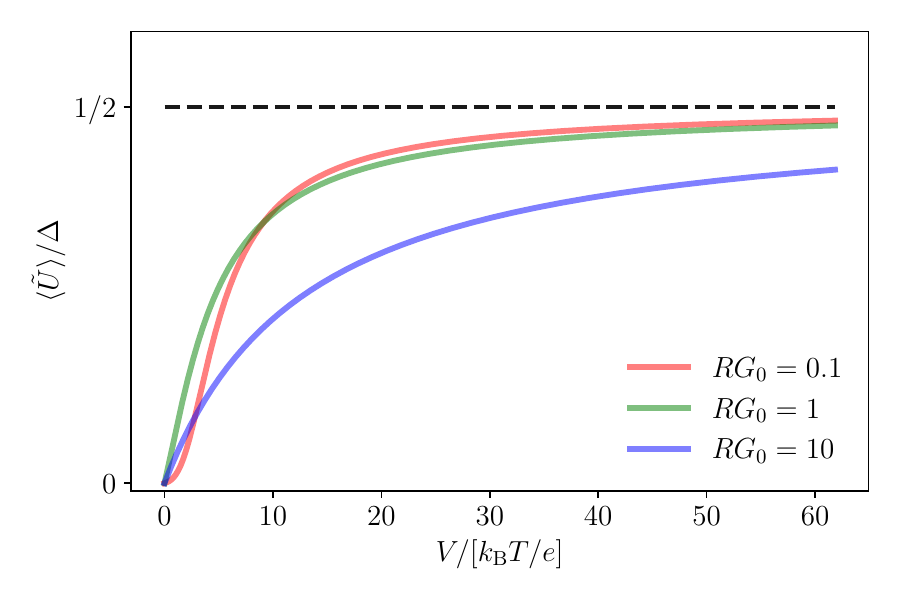}
    \caption{Rescaled environmental corrections for the DC voltage in a diode for various values of $RG$ at room temperature. For large positive values of the bias $V$, the corrections to the average voltage approach $\Delta/2$.}
    \label{fig:Diode}
\end{figure}

\paragraph*{Diode.} A diode is a device with nonlinear I-V characteristic, $ \mathcal{I}(u) = I_{\mathrm{D}} \big( \ee^{e u/\kb T} - 1 \big) $, but with noise also given by \eq \eqref{eq:shot_noise}. Here $ I_{\mathrm{D}} > 0 $ is maximum reverse current in the diode. Applying our general result \eq \eqref{eq:U_average} gives for the DC voltage (with $ T = T_{\textsc{r}} $):
\begin{eqal}
    \langle \tilde{U} \rangle & = \dfrac{\Delta R_\parallel G }{2} \Big( 1 - R_\parallel/R_{0, \parallel}  \Big),
\end{eqal}
where $G=\mathcal{I}'(u_0)=G_0 \ee^{e u_0/\kb T} $ is the conductance, $ u_0 $ is the solution of \eq \eqref{IdealIV}, $G_0=e I_{\mathrm{D}}/(\kb T)$ the conductance at equilibrium and $ R_{0, \parallel} = R/(1+RG_0) $ is the equilibrium parallel resistance. The environmental corrections to the average voltage are illustrated in \fig \ref{fig:Diode}. At large positive voltage bias $V$, one gets
\begin{eqal}
    \langle \tilde{U}\rangle & = \Delta/2.
\end{eqal}
Similarly to the tunnel junction, the correction reaches a limit related to the Coulomb gap $ \Delta $. The extra $ 1/2 $ factor comes from the rectification partially compensating the feedback.

\paragraph*{Conclusion}
We have provided a theoretical framework that enables the calculation of environmental corrections to all cumulants of voltage fluctuations across an arbitrary nonlinear, noisy component. We have dealt with an environment made of an RC circuit, but any passive network could be implemented in a similar way. We have shown how our results have relevant interpretation in terms of thermodynamics, in particular its link with Brillouin's paradox. Our predictions are within the reach of current experimental capabilities. Preliminary results were obtained \cite{Karan17} and quantitative measurements are in progress. Our approach opens a path toward the study of feedback effects in classical electronics. Next interesting steps will be to study multi-times correlations functions and response to an AC field: AC conductance, frequency-dependent noise, noise susceptibility, photo-assisted noise, etc.

\paragraph{Aknowledgements.}
We are very grateful to F. Bonani, M. Bonnin, J.-C. Delvenne, A. Dumont and L. van Brandt for fruitful discussions. This work was supported by the Canada Research Chair program, the NSERC, the Canada First Research Excellence Fund, the FRQNT, and the Canada Foundation for Innovation.

\bibliography{Biblio.bib}

\end{document}